# Identification of Two Regimes of Carrier Thermalization in PbS Nanocrystal Assemblies

*Augustin Caillas[1], Stéphan Suffit[1], Pascal Filloux[1], Emmanuel Lhuillier[2], Aloyse Degiron[1,*]*

[1]Université de Paris, CNRS, Laboratoire Matériaux et Phénomènes Quantiques, F-75205 Paris, France

[2]Sorbonne Université, CNRS, Institut des NanoSciences de Paris, INSP, F-75005 Paris, France

**Corresponding Author**

*E-mail: (A.D.) aloyse.degiron@u-paris.fr

ABSTRACT.

We bring fresh insight into the ensemble properties of PbS colloidal quantum dots with a critical review of the literature on semiconductors followed by systematic comparisons between steady-state photocurrent and photoluminescence measurements. Our experiments, performed with sufficiently low powers to neglect nonlinear effects, indicate that the photoluminescence spectra have no other noticeable contribution beside the radiative recombination of thermalized photocarriers (i.e. photocarriers in thermodynamic quasi-equilibrium). A phenomenological model based on the local Kirchhoff law is proposed that makes it possible to identify the nature of the thermalized photocarriers and to extract their temperatures from the measurements. Two regimes are observed: for highly compact assemblies of PbS quantum dots stripped from organic ligands, the thermalization concerns photocarriers distributed over a wide energy range. With PbS quantum dots cross-linked with 1,2-ethanedithiol or longer organic ligand chains, the thermalization solely concerns the fundamental exciton and can quantitatively explain all the observations, including the precise Stokes shift between the absorbance and luminescence maxima.

The connections between the luminescence and absorption properties of bulk semiconductors have been abundantly discussed in the literature[1–3]. Most aspects of this interdependence are well captured by the generalization of the Kirchhoff law for nonthermal radiation[1]—a model built upon the hypothesis that the excited electrons and holes at the origin of the luminescence are in thermodynamic equilibrium in the conduction and valence bands, respectively. Because of this equilibrium, the reservoirs of excited carriers can be described by intensive thermodynamic parameters (a temperature T, a chemical potential of radiation µ) and their energy distributions in the conduction and the valence bands follows Fermi-Dirac statistics. The Kirchhoff law is built on these hypotheses. It states that the spontaneous emission spectrum $I(\lambda)$ of bulk semiconductors is proportional to the product of their absorption cross-section $A(\lambda)$ times Planck's law, which, in most cases of interest, can be further simplified as:

$$I(\lambda) \propto A(\lambda).W(\lambda,T). \qquad (1)$$

In this relation, $W(\lambda,T)=2hc^2\lambda^{-5}\exp[-hc/(\lambda k_B T)]$ is the Wien approximation of Planck's law, h and $k_B$ are the Planck and Boltzmann constants and λ is the photon wavelength[1]. Although one can found various formulations of the Kirchhoff law in the semiconductor literature since at least the 1950s[2], the first fully consistent derivation seems to have been performed by Würfel in 1982[1]. The great insight of Würfel has been to realize that in addition to the carrier temperature T, it was necessary to introduce the chemical potential of radiation µ to make the model physically valid in all experimental situations. For simple cases, such as the spontaneous emission of photons under weak pumping power, µ simply appears in the proportionality factor between $I(\lambda)$ and the product $A(\lambda).W(\lambda,T)$ in eq 1. The model was later refined to encompass the quasi-equilibrium situations that arise when the excitation process creates populations of electrons and holes that are hotter than the crystal host. This problem is far more complex because the temperatures of the electron and hole reservoirs are not necessarily equal; nevertheless, it is still possible to describe such systems with the same formalism, except that T must be understood as an effective radiation temperature rather than the actual carrier temperature[4].

As a corollary, a semiconductor obeying the Kirchhoff law implies that its excited carriers have reached a global thermodynamic (quasi-) equilibrium. These electrons and holes, which we will henceforth denote as thermalized carriers when they are at (quasi-) equilibrium, fill the valence and conduction bands at all energies according to the product of a Fermi-Dirac statistic times the density of states. In practice, this regime is almost always reached for bulk semiconductors regardless of the way their carriers are excited (by heat, light, or electricity), even in time-resolved experiments where electrons and holes are initially excited with intense energetic laser pulses because the thermalization processes leading to the equilibrium typically occur over time scales smaller than a hundred of femtoseconds[3].

In the 1990s and early 2000s, the thermalization arguments behind the Kirchhoff law have also been invoked to explain the ensemble properties of more complex systems, such as quantum well heterostructures[5,6] and compact assemblies of epitaxial quantum dots[7,8]. Of central interest at that time was to elucidate the origin of the Stokes energy shift between the resonant absorption and luminescence peaks in these excitonic systems. The Kirchhoff law—or more exactly, a simplified version in which the Wien term is reduced to the Boltzmann distribution $\exp[hc/(\lambda k_B T)]$—provided the relevant framework to understand this shift, which turned out to be a direct consequence of the thermalization of the inhomogeneously broadened excitons. As originally pointed out by Gurioli et al.[6], if exciton thermalization does occur, then the Stokes energy shift must be equal to $\sigma^2/kT$, where $\sigma^2$ is the variance of the inhomogenously broadened quantum wells or quantum dots. This dependence can be mathematically derived from eq 1 by noting that the absorption of such systems is a sum of two terms: a resonant excitonic part $A_{RES}$, which can be approximated by a Gaussian distribution (in energy) with variance $\sigma^2$, and a non-resonant part $A_{NR}$. The $\sigma^2/kT$ prediction has proven strikingly accurate for a variety of epitaxial heterostructures with widely different inhomogeneous broadenings[6–8], proving, since this expression is derived from eq 1, that the excitons were thermalized in these experiments.

With the ongoing developments of solution-processed devices based on colloidal quantum dots (CQDs)[9–11] and nanoplatelets[12–14], a few studies[8,15–19] have also investigated the possibility of thermalization within such systems. However, it appears that a clear identification of which carriers thermalize in these structures is still lacking, some studies assuming that the thermalization concerns free carriers excited well above the first excitonic transition, while others consider that an equilibrium is reached for all electron-hole pairs, from the first excitonic resonance to all the other carriers excited at higher energy. As we shall see in the present study, not only are these assumptions not necessarily valid for PbS CQDs, but identifying which carriers are thermalized is crucial for correctly interpreting the mechanisms connecting the absorbance and luminescence in these composite assemblies.

In retrospect, one can wonder why the thermalization model has proven so successful and popular beyond bulk semiconductors because the electrons and holes that are involved in the thermalization within quantum well heterostructures and quantum dots are not merely free carriers characterized by a Fermi-Dirac distribution, but also, for a large part, excitons. Perhaps even more problematic is that the fundamental hypothesis upon which is built the Kirchhoff law is not met—namely, it is not possible to define a global thermodynamic equilibrium in these structures, implying, among others, that the chemical potential of radiation µ varies with the photon wavelength. In fact, most systems other than bulk semiconductors are always in an out-of-equilibrium situation, including homogeneous semiconducting layers if their thickness is smaller than their absorption depth[1]. In 2018, Greffet *et al.* solved this paradox by proving mathematically that it is still possible to define a local equilibrium for semiconductors plunged in inhomogeneous environments (i.e. thin films, nanostructures, and so on…)[20]. In other words, the carriers can still be treated as thermalized reservoirs, except that the thermodynamic parameters dictating their energy distributions (T and µ) are local quantities that are allowed to vary in space and time.

The first experimental validation of the local Kirchhoff law was obtained with assemblies of PbS quantum dots weakly coupled to plasmonic antennas[21]. The proof for carrier thermalization was provided, as many of the aforementioned studies, by showing that the antenna-enhanced luminescence was proportional to the Boltzmann statistics $\exp[hc/(\lambda k_B T)]$. In this regime, the role of the antennas can be understood as locally enhancing the absorption cross-section $A(\lambda)$ of the quantum dots due to their high local fields—which is equivalent to increasing the luminescence $I(\lambda)$ at the same wavelength $\lambda$. An important consequence of these findings is that the Purcell effect[22], which has been so far almost universally invoked in the context of optical antennas, is not necessarily relevant for emitters experiencing carrier thermalization. Contrarily to the Purcell effect, the local Kirchhoff law does not involve a competition between radiative and non-radiative decay channels. It provides a satisfying explanation of the main results reported in Ref.[21]: (i) the fact that the emitters were not quenched when placed in direct contact with the lossy antennas and (ii) the fact that the antennas were capable of triggering light emission at wavelengths significantly different from the bandgap of the quantum dots because the thermalized carriers occupy energy states also above the band edge.

However, this experimental study left several points unanswered. Chief among them, and just as in the past literature on PbS CQDs layers without antennas briefly evoked above, it did not identify which carriers where thermalized. The present article aims at answering this question and, more generally, at providing a more complete and self-consistent description of the ensemble properties of PbS CQDs. For more generality, we will consider samples without antennas so as to identify and analyze properties that are not restricted to special cases dictated by the local environment.

Our study is based on systematic comparisons between the absorption and photoluminescence spectra of compact layers of PbS CQDs in the steady state. As shown on **Figure 1**a, the samples consist of a layer of CQDs spun on a glass substrate with Au interdigitated electrodes. The electrodes are used to establish the absorption spectrum $A(\lambda)$ of the film via Fourier transform photocurrent spectroscopy. Admittedly, this technique is less straightforward than estimating the absorbance from transmittance and reflectance spectra of samples without electrodes (since absorbance = 1 – transmittance – reflectance). However, it is more accurate for the thin layers that we consider (see below) because it does not suffer from the same artifacts such as light trapped within the substrate and diffuse scattering that escapes optical detection. The structures are biased at 11 V and illuminated with a 10X objective using an incandescent tungsten light source passing through a Fourier Transform Infrared (FTIR) spectrometer (Invenio-R from Bruker). The photocurrent that flows between the electrodes is amplified, converted into a voltage and plugged back into the FTIR electronics. By normalizing the data with the spectrum measured with a calibrated 818-ST2 Ge photodiode from Newport, one obtains the responsivity of the sample which is itself proportional to the absorbance of the CQD film after dividing the data by the wavelength[23]. We have verified that the spectra do not depend on the spacing between the interdigitated electrodes or the applied voltage (**Figures S1**a and S1b). We have also ensured that the data do not change if one uses the 50X objective of the PL experiments described in the next paragraph (Figure S1c).

To record a PL spectrum I($\lambda$), we pump the sample with a continuous HeNe laser at 633 nm. This laser is focused with a 50X microscope objective that is also used to collect the infrared PL. A dichroic mirror is used to separate the IR signal from the visible pump. The spectrum is measured using an Acton SP2356 grating spectrometer coupled to a NIRvana InGaAs camera. The measurements are performed between the interdigitated electrodes, on the CQDs that are directly in contact with the glass substrate.

The films of PbS CQDs considered in this study are only 15 to 20 nm thick depending on the ligand cross-linking procedure used to adjust their compactness, mitigating artificial blue shifts of the PL due to optical reabsorption.[24] While such thickness is much smaller than that of the interdigitated electrodes (70 nm), we have ensured via scanning electron microscope observations that the CQDs layer was uniform except for a small accumulation in the immediate vicinity of the electrodes (**Figure S2**). The samples are characterized in air without encapsulation.

We start our study with a film of PbS CQDs cross-linked with a $Na_2S$ solution, resulting in a highly compact layer where the individual CQDs are separated by $S^{2-}$ dangling ions. The sample is n-doped (ie presents a rise of conductance under positive gate bias), as can be judged from the transistor measurements of **Figure S3** performed with an electrolytic gate above the film.[25] Figure 1b shows on the same plot the absorption A($\lambda$) of the sample together with the PL spectra I($\lambda$) recorded for increasing pumping powers $P_1$ = 5.5 µW, $P_2$ = 22 µW, $P_3$ = 138 µW and $P_4$ = 550 µW (here and throughout the study, the corresponding power densities in W/cm$^2$ are roughly 30 times these values with the 50X objective used to focus the HeNe laser onto the sample). For better visibility, the PL data are normalized between 0 and 1 (unnormalized data appears on **Figure S4a**). The results are consistent with the literature[21,26]—the absorption spectrum features a resonant shoulder at $\lambda = 1280$ nm that emerges from a non-resonant tail. The shoulder corresponds to the excitation of the first exciton. Its spectral signature is weak because the extreme compactness of the Na$_2$S-treated film relaxes the quantum confinement within the PbS CQDs (in contrast, the absorption spectrum of the PbS CQDs with their native oleic acid ligands in solution, displayed in **Figure S5**, features a sharp excitonic maximum). The PL spectra, for their parts, are characterized by an inhomogeneously broadened peak with a sizeable Stokes shift compared to the position of the excitonic absorption shoulder. Moreover, this peak exhibits a minor shift toward smaller wavelengths as the pumping power increases. The evolution of the PL spectrum as a function of the pump power is also represented in a more systematic way on Figure 1c in a two-dimensional color plot that contains data for 11 pumping powers ranging from 5.5 to 550 µW.

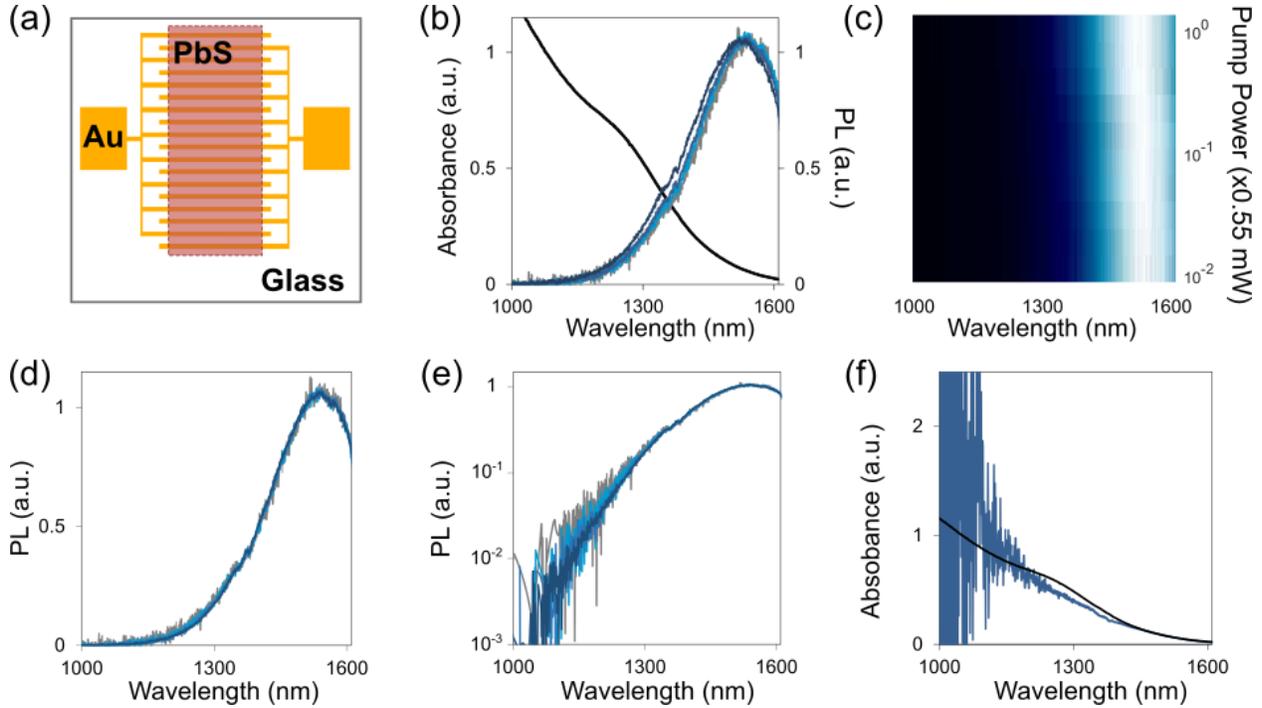

**Figure 1.** Properties of the layer of PbS CQD cross-linked with $Na_2S$. (a) Schematic top view of a sample. (b) Absorbance spectrum (black curve) and PL spectra normalized between 0 and 1 for increasing pump powers (5.5 µW for the grey curve, 22 µW, 138 µW and 550 µW, for the light, medium and dark blue curves, respectively). The oscillations around 1360 nm are artifacts corresponding to absorption lines of the atmosphere. (c) PL intensity (normalized between 0, black and 1, white) as a function of the wavelength (horizontal axis) and pump power (vertical axis in log scale). This plot has been created with 11 PL spectra, taken with pump powers ranging from 5.5 µW to 550 µW, with a logarithmic increment of 0.2 between each subsequent measurement. (d) Result of the Boltzmann renormalization described in the text in linear scale. (e) Result of the Boltzmann renormalization in logarithmic scale. (f) Absorbance of the CQD film [black curve, same as panel (a)] compared to the $I_4(\lambda)/W(\lambda,T_4)$ ratio obtained with a pump power of 550 µW and a retrieved temperature $T_4 = 400$ K (blue curve). The noise on the left side is caused by the vanishing PL at small wavelengths.

We now show that these features are those of a system governed by thermalized resonant and non-resonant photocarriers. As discussed above, the thermalization hypothesis entails that the different PL spectra of Figure 2b are proportional to $\exp[-hc/(\lambda k_B T)]$. As a consequence, it should be possible to find a set of temperatures $T_i$, corresponding to the different pumping powers $P_i$, that makes it possible to transform any of these spectra into the one recorded with the lowest pumping power $P_1$ by multiplying the data by $\exp[hc/(\lambda k_B T_i)]/\exp[hc/(\lambda k_B T_1)]$. Figures 1d,e prove, both in linear and logarithmic scale, that such temperatures can be found if one chooses $T_1$= 368 K, $T_2$ = 372 K, $T_3$ = 381 K and $T_4$= 400 K.

The results of Figures 1d,e unambiguously prove that the system is thermalized. To go deeper into the analysis, we test the hypothesis of eq 1, which predicts that $A(\lambda)$ can be recovered from any of the PL spectra $I_i(\lambda)$ by dividing them by $W(\lambda,T_i)=2hc^2\lambda^{-5}\exp[hc/(\lambda k_B T_i)]$, with $\{T_i\}_{i=1,2,3,4}$ the temperatures found with the renormalization procedure shown in Figures 1d,e.

Figure 1f shows that the ratio $I_4(\lambda)/W(\lambda,T_4)$ overlaps well with $A(\lambda)$—except for the excitonic shoulder that is not fully reconstructed. The same result can be obtained with any of the other ratios $I_i(\lambda)/W(\lambda,T_i)$. In other words, it appears that eq 1 is verified—except that the excitonic contribution is weaker than expected. This conclusion is fully consistent with the fact that excitons in dense packs of PbS CQDs tend to dissociate spontaneously to generate free carriers[27], leading to a significant quenching of the PL as a result. Thus, the results of Figure 1 imply that the optoelectronics properties of the PbS layer treated with $Na_2S$ result from the thermalization of carriers that do not merely accumulate at the band edge, but also populate energy levels at higher energies. This claim is further supported by the first experimental study on the local Kirchhoff law that we have discussed in the introduction[21]: by coupling a film of PbS CQDs treated with $Na_2S$ with metallic antennas, it was possible to trigger radiative transitions at energies significantly higher than the bandgap, which would not have been possible if the excited carriers were accumulating at the band edge only.

Because the values of $T_i$ are larger than the room temperature, we are dealing with a quasi-equilibrium situation, which, as noted in the introduction, raises the possibility that the thermalized electrons and holes do not share the same temperature. In this context, T must be understood as an effective radiation temperature[4] rather than the actual electron and hole temperatures.

We now repeat the same experiments with PbS CQDs cross-linked with 1,2-Ethanedithiol (EDT), resulting a slightly less compact emitter assembly (with EDT, the distance between adjacent CQDs is approximately 0.25 nm). Field effect transistor measurements indicate that the film is also n-doped (Figure S3b) and that its carrier mobility is roughly six times smaller than for the previous case (see Experimental Methods). This observation, together with a much brighter emission (Figure 2a), are expected and result from an improved quantum confinement of the electron wavefunctions when the inter-dot distance increases even by a few ångströms[28]. The absorbance spectrum and a series of PL spectra obtained with different pump powers are presented on Figures 2b,c. At first, it is tempting to conclude that this second example is similar to the $Na_2S$ case, except that all the trends are more pronounced. In particular, the excitonic peak in the absorbance spectrum is now well defined because the larger inter-dot spacing improves the quantum confinement of the electron wavefunctions. Moreover, the shift of the PL as a function of the pumping power is more dramatic even though Figures 3d,e show that the different PL spectra can still be almost perfectly superimposed by renormalizing them with a ratio of two Boltzmann distributions. Said otherwise, the system is also thermalized.

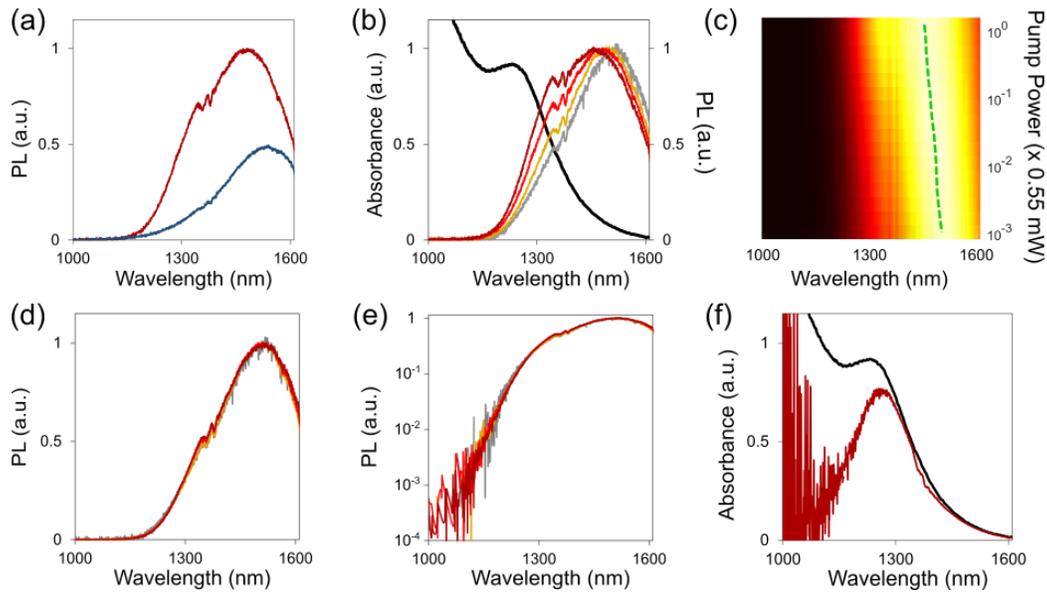

**Figure 2.** Properties of the layer of PbS CQD cross-linked with EDT. The oscillations and/or PL shoulder around 1360 nm that are visible on most panels are artifacts in the PL spectra corresponding to absorption lines of the atmosphere. (a) PL of the film treated with EDT (red curve) and the film treated with the $Na_2S$ (blue curve). Both spectra have been recorded with a pump power of 55 µW. The relative intensity between the two curves is the actual one. (b) Absorbance spectrum (black curve) and PL spectra for increasing pump powers (0.55 µW for the grey curve, 5.5 µW for the yellow curve, 55 µW and 550 µW for the light and dark red traces, respectively). (c) PL intensity (normalized between 0, black and 1, white) as a function of the wavelength (horizontal axis) and pump power (vertical axis in log scale). This plot has been created with 16 PL spectra, taken with pump powers ranging from 0.55 µW to 550 µW, with a power increment of 0.2 between each subsequent measurement. The green dashed curve is the predicted position of the PL maximum knowing the position of the absorption peak, its broadening $\sigma^2$ and the exciton temperatures retrieved from the Boltzmann renormalization described in the text. (d) Result of the Boltzmann renormalization described in the text in linear scale. (e) Result of the Boltzmann renormalization in logarithmic scale. (f) Absorbance of the nanocrystal film [black curve, same as panel (a)] compared to the $I(\lambda)/W(\lambda,T)$ ratio obtained with a pump power of 550 µW and a retrieved temperature T = 439 K (red curve). The noise on the left side is caused by the vanishing PL at small wavelengths.

However, an examination of the $I(\lambda)/W(\lambda,T)$ ratios reveals crucial differences compared to the Na$_2$S case. Figure 2f shows that these ratios are actually only overlapping with the resonant part of the absorption $A_{RES}(\lambda)$, corresponding to the creation of excitons. This observation implies that the PL emission is essentially governed by the thermalization of excitons:

$$I(\lambda) \propto A_{RES}(\lambda).W(\lambda,T), \qquad (2)$$

with T the temperature extracted from the renormalization procedure yielding Figures 2c-d. All the other electron-hole pairs (multi-excitons, free carriers…), which are responsible for the non-resonant absorption tail at smaller wavelengths, do not seem to have a sufficient lifetime to thermalize. This conclusion is also consistent with the fact that electrical conduction within ensembles of PbS CQDs treated with EDT is ensured by phonon-assisted hopping rather than a band-like transport involving free carriers[29]. In other words, this situation is at variance with that of the PbS film treated with Na$_2$S for which carriers at energies higher than the excitonic peaks were also thermalized. In addition, contrarily to the Na$_2$S case, there is no difficulty about the meaning of the temperature T—the latter characterizes the excitons, which are the only particles that thermalize.

To confirm these claims, we next examine how the different PL maxima of Figure 2 are related to the position of the excitonic absorption peak. As explained in the introduction, an unambiguous signature of exciton thermalization is the $\sigma^2/kT$ dependence of the Stokes energy shift, where $\sigma^2$ is the variance of the inhomogeneous broadening approximated as a Gaussian distribution[6]. Given the wavelength of the absorption peak, its full width at half maximum revealed by Figure 2f, and the temperatures extracted from the renormalization procedure, we can reconstruct the expected positions of the PL maxima for each pumping power. These positions are represented as a dashed green line on the two-dimensional PL plot of Figure 2c which contains 16 independent measurements corresponding to pumping powers varying over three orders of magnitude, from 0.55 µW to 550 µW. They correctly predict the experimental trends, providing a self-consistent validation of the exciton thermalization model, the retrieved exciton inhomogeneous broadening $\sigma^2$ and their temperature T. We note for the sake of completeness that the $\sigma^2/kT$ dependence is actually an approximation of the exact Stokes shift. As noted in the introduction, this expression has been originally introduced in the context of epitaxial heterostructures using a Kirchhoff law relation in which the Wien term is reduced to a simple Boltzmann distribution[6]. It can be readily seen that such an approximation, which neglects the $\lambda^{-5}$ dependence of the Wien term, induces a minor overestimation of the actual Stokes shift (see eq S1 for the exact expression).

The fact that exciton thermalization plays a central role in the photoluminescence of PbS CQDs treated with EDT is further confirmed by having a closer look at the retrieved temperatures. These temperatures, plotted on **Figure 3**a, logarithmically raise from 363 K to 439 K as the pumping power increases from 0.55 µW to 550 µW. They are significantly larger than the lattice (room) temperature and can only be explained if the thermalized electron-hole pairs are strongly confined (as opposed to free carriers). Moreover, these temperatures have been obtained after letting the CQD layer oxidize several days. The temperatures were significantly lower when the sample was characterized hours after fabrication (light red data set of Figure 3a), which we interpret as a consequence of a lower quantum confinement due to a thinner oxide shell around the PbS CQDs. This hypothesis is supported by **Figure S6** which shows that the absorption and PL peaks of the fresh sample were initially observed at longer wavelengths, indicating that the radius of the PbS core was originally larger (i.e. the oxide shell was either not yet formed or incomplete or much thinner than after several days of exposure in air).

The other data sets on Figure 3a are the retrieved temperatures for the sample treated with $Na_2S$, extracted days (blue curve) and hours (light curve) after its fabrication. The same logarithmic dependence as for the EDT sample is observed but the temperatures are lower and tending to room temperature for the lowest pumping powers in the non-oxidized case. This observation provides a further validation of the interpretations put forward in this study, as the quantum confinement within the PbS film treated with $Na_2S$ is relaxed due to the reduced inter-dot spacing.

The behavior elucidated in this article represents a marked departure from bulk semiconductors which are characterized by a Fermi-Dirac distribution of thermalized free carriers. It is also different from the case of quantum well heterostructures and epitaxial quantum dots, for which exciton thermalization associated with a $\sigma^2/kT$ Stokes shift was originally demonstrated[6–8]. In these studies, the thermalization involved the full absorption spectrum $A(\lambda) = A_{RES}(\lambda) + A_{NR}(\lambda)$, the latter term including the participation of multi-excitons and free carriers at high energy. We justify this difference by the fact that our CQDs are plunged in an environment characterized by voids, organic residues, surface traps and energy barriers of various heights, as opposed to ordered lattices for epitaxial quantum wells or epitaxial quantum dots.

In this respect, one can wonder about the generality of our findings, and, in particular, whether the exciton thermalization observed for the sample treated with EDT also holds for PbS CQDs capped with longer ligand chains. To answer this question, we show on Figures 3b-3f the results obtained with PbS CQDs that have conserved their native, 2-nm-long, acid oleic ligands. Only the PL data are plotted because we were not able to measure a reliable absorbance spectrum with such long insulating ligand chains (an estimation of the absorbance, obtained with optical measurements rather than photocurrent experiments, is plotted on Figure S7a and compared with the $I(\lambda)/W(\lambda,T)$ ratios of Figure 3f). The same general features as in Figure 2 are observed, implying that the exciton thermalization is robust indeed. However, the data also reveal deviations from this mechanism because the Boltzmann renormalization of Figures 3d-3e is not as perfect as in the EDT case. Although a detailed investigation of these deviations is out of the scope of this study, they suggest that other processes compete with the thermalization for PbS CQDs capped with long ligand chains. On one hand, the small deviations observed on Figures 3d-3e induces a large uncertainty on the retrieved temperatures (plotted in Figure S7b with error bars) because different portions of the curves can be used for the renormalization procedure. On the other hand, the thermalization of excitons remains the dominant process because the $I(\lambda)/W(\lambda,T)$ ratios have the same excitonic-like shape at all pumping powers, as evidenced on Figure 3f for the weakest and strongest laser illumination considered in this study.

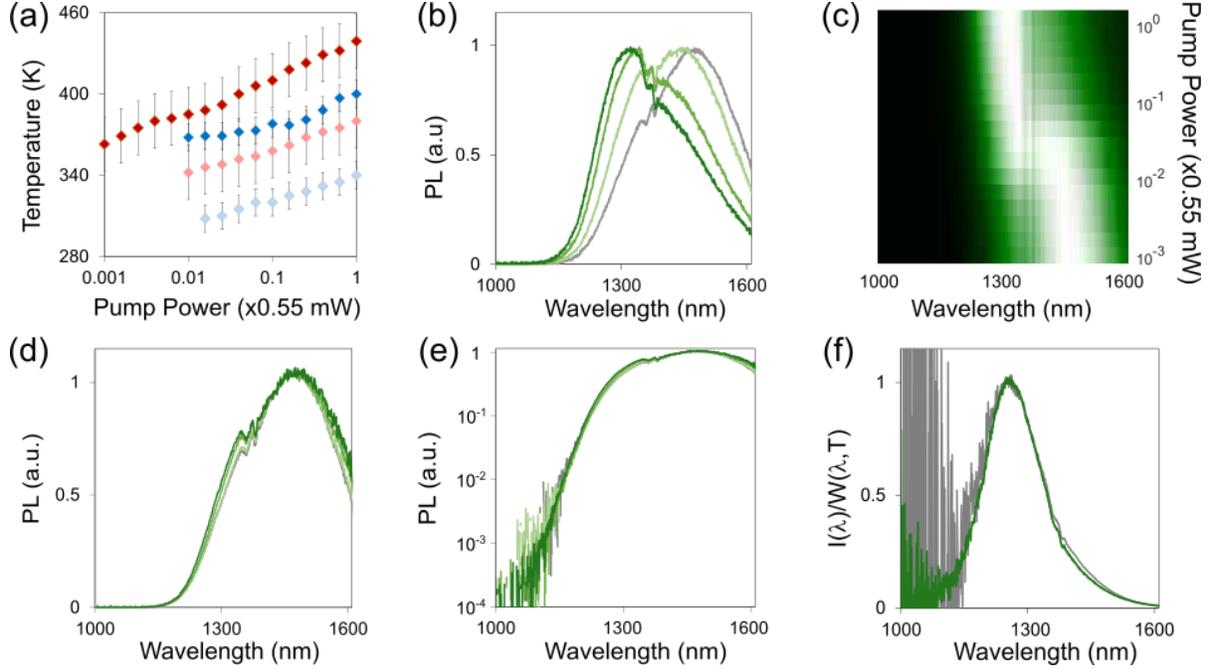

**Figure 3.** (a) Summary of the retrieved temperatures for the two CQD layers considered previously. Light red: fresh EDT-treated film. Darker red: EDT-treated film after several days weeks in air. Light blue: fresh Na$_2$S treated film. Darker blue: Na$_2$S treated film after several days in air. (b) PL spectra of a PbS CQD film with native acid oleic ligands for increasing pump powers (0.55 µW for the grey curve, 5.5 µW for the yellow curve, 55 µW and 550 µW for the light and dark red traces, respectively). (c) PL intensity for the PbS CQD film with native acid oleic ligands (normalized between 0, black and 1, white) as a function of the wavelength (horizontal axis) and pump power (vertical axis in log scale). This plot has been created with 16 PL spectra, taken with pump powers ranging from 0.55 µW to 550 µW, with a power increment of 0.2 between each subsequent measurement. (d) Result of the Boltzmann renormalization described in the text in linear scale. (e) Result of the Boltzmann renormalization in logarithmic scale. (f) $I(\lambda)/W(\lambda,T)$ ratios obtained with a pump power of 550 µW and a retrieved temperature T = 580 K (green curve) and with a pump power of 0.55 µW and a retrieved temperature T = 375 K (gray curve). The noise on the left side is caused by the vanishing PL at small wavelengths. The oscillations and/or PL shoulder around 1360 nm that can be seen on most panels are artifacts in the PL spectra corresponding to absorption lines of the atmosphere.

In conclusion, our experiments show that the optoelectronic properties of ensembles of PbS CQDs in the steady state regime are essentially governed by the thermalization of photo-excited carriers, where the thermalization must be understood as a local thermodynamic equilibrium. Moreover, the nature of thermalized carriers evolves from electrons and holes distributed over a large energy range (in the limit case of PbS CQDs stripped from organic ligands) to confined excitons centered on a much narrower spectral window. As a corollary, the exciton temperature can be tuned from the lattice temperature to several more hundreds of kelvins by adjusting the pump power and the size of the ligands capping the semiconducting CQDs. Our measurements suggest that previous hypotheses made to extract the carrier temperature within layers of PbS CQDs may need reconsideration, some past studies focusing on the thermalization of free carriers at high energy[18], while others relied on fitting the PL spectra with a sigmoid function that both encompasses the resonant absorption and the high-energy absorption tail[16,17]. It remains to be seen how these conclusions evolve under much higher pumping powers, when the Wien approximation of Planck's law is not valid anymore and optical nonlinearities must be taken into account, such as the fact that the absorption cross-section of the emitters become strongly excitation-dependent. Likewise, it would be interesting to see how much thermalization remains for PbS CQDs in solution, since the sizeable Stokes shifts that are also observed in such cases are also a consequence of interdot interactions[24]. Last, the same methodology could be applied to other types of emitters, and to more complex systems that involve optical antennas or other tailored photonic environments.

EXPERIMENTAL METHODS

**Chemicals.** Poly(ethylene glycol) (PEG, Sigma Aldrich, molecular weight Mn = 6000 g.mol$^{-1}$), LiClO$_4$ (Sigma Aldrich, 99.99%), oleylamine (OLA, Acros, 80-90%), %), lead chloride (PbCl$_2$, Afla Aesar, 99%), sulfur powder (S, Afla Aesar, 99.5%), oleic acid (OA, Afla Aesar, 90%), trioctylphosphine (TOP, Afla Aesar, 90%),n-hexane (VWR), Ethanol (VWR, >99.9%), Toluene (Carlo Erba, >99.8%).

**Precursor.** Lead oleate Pb(OA)$_2$ 0.1 M: 0.9 g of PbO are mixed in a three neck flask with 40 mL of oleic acid. The flask is degassed under vacuum for one hour. The atmosphere is switched to Ar and the temperature raised to 150 °C for two hours. The final solution is typically clear and yellowish.

**PbS CQD Synthesis.** In a three-neck flask, 300 mg of PbCl$_2$, together with 100 µL of TOP and 7.5 mL of OLA are degassed, first at room temperature and then at 110 °C for 30 min. Meanwhile, 30 mg of S powder is mixed with 7.5 mL of OLA until full dissolution and an orange clear solution is obtained. Then under nitrogen at 80 °C, this solution of S is quickly added to the flask. After 2 minutes, the reaction is quickly quenched by addition of 1 mL of OA and 9 mL of hexane. The nanocrystals are precipitated with ethanol and redispersed in 5 mL of toluene. This washing step is repeated one more time and the pellet is this time dispersed in 10 mL of toluene with a drop of OA. The solution is then centrifugated at is, to remove the unstable phase. The supernatant is precipitated with methanol and redispersed in toluene. Finally, the PbS CQD solution in toluene is filtered through a 0.2 µm PTFE filter. The obtained solution is used for further characterization and devices' fabrication.

**Sample fabrication.** The interdigitated electrodes are defined by optical lithography on a borosilicate glass substrate from Plan Optik using an MJB4 aligner and the AZ 5214E photoresist. After deposition of 3 nm of Ti and 70 nm of Au in a Plassys MEB 550S electron beam evaporator, followed by a lift-off in acetone, the resulting electrodes define an area of $870 \times 1400$ µm$^2$ and the spacing between two consecutive digits is 10 µm. The remaining steps of the fabrication are performed in an $N_2$-filled glovebox. A 15-20 nm thick layer of CQDs is spun onto the sample, creating a coating above the metallic patterns. An examination of various cross-sections with a scanning electron microscope reveal that the thickness of the layer is uniform between the interdigitated electrodes, except for small edge effects in their immediate vicinity. Finally, the sample is dipped into a solution of EDT in ethanol for 60 s or into a solution of $Na_2S$ in ethanol for 10s followed by a rinse in ethanol to replace the native OA ligands by shorter molecules. For the transistor measurements, an additional step[30] is necessary to fabricate and deposit the electrolytic gel onto the layer of CQDs. The latter is obtained by mixing 230 mg of PEG and 50 mg of $LiClO_4$ at 150 °C for several hours in the glovebox. The resulting mixture is then cooled down until it solidifies and then reheated at 100 °C for 10 minutes to soften it prior deposition onto the sample.

**Transistor measurements.** The characterization is performed under ambient atmosphere with an electrolytic gate applied above the CQD layer. Automatic transfer curves are measured using a Keithley 2636B sourcemeter. The source and drain are the interdigitated electrodes and the field effect is applied by contacting an electrical probe to the electrolytic gate[30]. The transistor characterizations are performed after all the other experiments because the latter cannot be done anymore once the electrolytic gate covers the CQDs. From these measurements, the carrier mobility of the samples can be obtained using the slope of the transfer curve:

$$\mu^{FET} = \frac{L}{WCV_{DS}} \frac{\partial I_{DS}}{\partial V_G}\bigg|_{V_{DS}},$$

where L is the electrode spacing (10 µm), W is the electrode length (43 x 1.4 mm), C is the gate surface capacitance and $V_{DS}$ is the drain source bias (0.25 V). $I_{DS}$ is the drain source current and $V_G$ is the gate bias.

**Absorption measurements.** The characterization is performed under ambient atmosphere using Fourier transform photocurrent spectroscopy. The samples are placed under a Hyperion microscope from Bruker equipped with a 10X infrared objective from Olympus, biased at 11 V (10.8 V to be more exact, which is the maximum value that we can reach with our apparatus) and illuminated with the white light originating from a Fourier Transform Infrared (FTIR) spectrometer (Invenio R from Bruker). The resulting photocurrent is amplified using a DLPCA-200 amplifier from Femto, converted into a voltage and plugged back into the electronics of the FTIR via an analog/digital converter. The spectra obtained with this technique are then normalized with the spectrum taken with a calibrated 818-ST2 Ge photodiode from Newport. At this stage we have the responsivity of the sample, from which we can derive the absorbance spectrum in arbitrary units by dividing the data by the wavelength[23]. Note that our structures are not good photodetectors because the thickness (and therefore the absorbance) of the CQDs is very small.

**Photoluminescence measurements.** The characterization is performed under ambient atmosphere. The sample is investigated with an Olympus BX51WI upright microscope equipped with an LCPLN50XIR 50X infrared objective. The laser pump is a HeNe laser at 633 nm, filtered with a BG40 filter from Thorlabs to remove the parasitic spontaneous emission in the near-infrared. The laser beam is directed through a set of neutral density filters that makes it possible to adjust its intensity, then directed into the microscope via the top port and finally focused onto the sample with the 50X objective. The infrared luminescence, collected by the same objective, is separated from the laser pump with a Thorlabs DMLP950R dichroic mirror and a RG780 long pass filter. The signal is analysed with an Acton SP-2356 spectrograph equipped with a 85 groove/mm monochromator and coupled to a NIRvana 640ST InGaAs camera from Princeton Instruments. The resulting data are then corrected by taking into account the transfer functions of the 50X objective and the 85 groove/mm grating. The dispersion of the other elements of the setup is sufficiently small to neglect their impact on the spectra.

**Why we use two different microscope objectives for the absorption and PL measurements.** We chose a 10X objective for the absorption measurements because the device cannot be illuminated in full with a higher magnification, resulting in a degraded signal-over-noise ratio (but the spectra are the same regardless of the objective used). We chose a 50X objective for the PL measurements for two reasons. First, this magnification ensures that the pumping spot is much smaller than the spacing between two interdigitated electrodes; second, this objective has the largest numerical aperture that we have at hand (0.65), maximizing the solid angle of collection and therefore the PL signal.

ASSOCIATED CONTENT.

The following files are available free of charge at

https://pubs.acs.org/doi/suppl/10.1021/acs.jpclett.1c01206/suppl_file/jz1c01206_si_001.pdf

Additional absorption spectra proving that the measurements are independent from the applied voltage, the spacing between the interdigitated electrodes and the objective used to focus light onto the sample (Figure S1); Scanning electron micrograph showing a cross-section of a device (Figure S2); Transistor measurements on the CQD films cross-linked with $Na_2S$ and EDT considered in the main text (Figure S3); PL spectra of Figures 1b and 2b without normalizing the data between 0 and 1 (Figure S4); Absorbance of the PbS CQDs in solution (Figure S5); Evolution of the absorbance and PL spectra with time (Figure S6); More results on the sample with native oleic acid ligands (Figure S7).

A.C. fabricated the samples with help of S.S. and P.F. The PbS CQDs were synthesized by E.L. who also prepared the electrolytic gel and shared his knowledge on ligand exchange, CQD processing and transistor measurements. The experiments and their analysis were performed by A.C. with the help of A.D. The manuscript was prepared by A.C. and A.D. with contributions of all authors. A.D. conceived the project.

The authors declare no competing financial interest

We acknowledge support from the European Research Council grant FORWARD (reference: 771688).

REFERENCES

(1) Würfel, P. The Chemical Potential of Radiation. *J. Phys. C Solid State Phys.* **1982**, *15* (18), 3967–3985. https://doi.org/10.1088/0022-3719/15/18/012.

(2) van Roosbroeck, W.; Shockley, W. Photon-Radiative Recombination of Electrons and Holes in Germanium. *Phys. Rev.* **1954**, *94* (6), 1558–1560. https://doi.org/10.1103/PhysRev.94.1558.

(3) Nozik, A. J. Spectroscopy and Hot Electron Relaxation Dynamics in Semiconductor Quantum Wells and Quantum Dots. *Annu. Rev. Phys. Chem.* **2001**, *52* (1), 193–231. https://doi.org/10.1146/annurev.physchem.52.1.193.

(4) Gibelli, F.; Lombez, L.; Guillemoles, J.-F. Two Carrier Temperatures Non-Equilibrium Generalized Planck Law for Semiconductors. *Phys. B Condens. Matter* **2016**, *498*, 7–14. https://doi.org/10.1016/j.physb.2016.06.006.

(5) Humlíček, J.; Schmidt, E.; Bočánek, L.; Švehla, R.; Ploog, K. Exciton Line Shapes of GaAs/AlAs Multiple Quantum Wells. *Phys. Rev. B* **1993**, *48* (8), 5241–5248. https://doi.org/10.1103/PhysRevB.48.5241.

(6) Gurioli, M.; Vinattieri, A.; Martinez-Pastor, J.; Colocci, M. Exciton Thermalization in Quantum-Well Structures. *Phys. Rev. B* **1994**, *50* (16), 11817–11826. https://doi.org/10.1103/PhysRevB.50.11817.

(7) Patanè, A.; Levin, A.; Polimeni, A.; Eaves, L.; Main, P. C.; Henini, M.; Hill, G. Carrier Thermalization within a Disordered Ensemble of Self-Assembled Quantum Dots. *Phys. Rev. B* **2000**, *62* (16), 11084–11088. https://doi.org/10.1103/PhysRevB.62.11084.

(8) Patanè, A.; Levin, A.; Polimeni, A.; Eaves, L.; Main, P. C.; Henini, M.; Hill, G. Universality of the Stokes Shift for a Disordered Ensemble of Quantum Dots. *Phys. Status Solidi B* **2001**, *224* (1), 41–45. https://doi.org/10.1002/1521-3951(200103)224:1<41::AID-PSSB41>3.0.CO;2-S.

(9) Weidman, M. C.; Yager, K. G.; Tisdale, W. A. Interparticle Spacing and Structural Ordering in Superlattice PbS Nanocrystal Solids Undergoing Ligand Exchange. *Chem. Mater.* **2015**, *27* (2), 474–482. https://doi.org/10.1021/cm503626s.

(10) Kagan, C. R.; Lifshitz, E.; Sargent, E. H.; Talapin, D. V. Building Devices from Colloidal Quantum Dots. *Science* **2016**, *353* (6302), aac5523–aac5523. https://doi.org/10.1126/science.aac5523.

(11) Gilmore, R. H.; Winslow, S. W.; Lee, E. M. Y.; Ashner, M. N.; Yager, K. G.; Willard, A. P.; Tisdale, W. A. Inverse Temperature Dependence of Charge Carrier Hopping in Quantum Dot Solids. *ACS Nano* **2018**, *12* (8), 7741–7749. https://doi.org/10.1021/acsnano.8b01643.

(12) Ithurria, S.; Tessier, M. D.; Mahler, B.; Lobo, R. P. S. M.; Dubertret, B.; Efros, Al. L. Colloidal Nanoplatelets with Two-Dimensional Electronic Structure. *Nat. Mater.* **2011**, *10* (12), 936–941. https://doi.org/10.1038/nmat3145.

(13) Polovitsyn, A.; Dang, Z.; Movilla, J. L.; Martín-García, B.; Khan, A. H.; Bertrand, G. H. V.; Brescia, R.; Moreels, I. Synthesis of Air-Stable CdSe/ZnS Core–Shell Nanoplatelets with Tunable Emission Wavelength. *Chem. Mater.* **2017**, *29* (13), 5671–5680. https://doi.org/10.1021/acs.chemmater.7b01513.

(14) Tenney, S. M.; Vilchez, V.; Sonnleitner, M. L.; Huang, C.; Friedman, H. C.; Shin, A. J.; Atallah, T. L.; Deshmukh, A. P.; Ithurria, S.; Caram, J. R. Mercury Chalcogenide Nanoplatelet–Quantum Dot Heterostructures as a New Class of Continuously Tunable

(14) Bright Shortwave Infrared Emitters. *J. Phys. Chem. Lett.* **2020**, *11* (9), 3473–3480. https://doi.org/10.1021/acs.jpclett.0c00958.
(15) Pelton, M.; Ithurria, S.; Schaller, R. D.; Dolzhnikov, D. S.; Talapin, D. V. Carrier Cooling in Colloidal Quantum Wells. *Nano Lett.* **2012**, *12* (12), 6158–6163. https://doi.org/10.1021/nl302986y.
(16) Cao, W.; Zhang, Z.; Patterson, R.; Lin, Y.; Wen, X.; Veetil, B. P.; Zhang, P.; Zhang, Q.; Shrestha, S.; Conibeer, G.; Huang, S. Quantification of Hot Carrier Thermalization in PbS Colloidal Quantum Dots by Power and Temperature Dependent Photoluminescence Spectroscopy. *RSC Adv.* **2016**, *6* (93), 90846–90855. https://doi.org/10.1039/C6RA20165B.
(17) Zhang, P.; Feng, Y.; Wen, X.; Cao, W.; Anthony, R.; Kortshagen, U.; Conibeer, G.; Huang, S. Generation of Hot Carrier Population in Colloidal Silicon Quantum Dots for High-Efficiency Photovoltaics. *Sol. Energy Mater. Sol. Cells* **2016**, *145*, 391–396. https://doi.org/10.1016/j.solmat.2015.11.002.
(18) Cao, W.; Yuan, L.; Patterson, R.; Wen, X.; Tapping, P. C.; Kee, T.; Veetil, B. P.; Zhang, P.; Zhang, Z.; Zhang, Q.; Reece, P.; Bremner, S.; Shrestha, S.; Conibeer, G.; Huang, S. Difference in Hot Carrier Cooling Rate between Langmuir–Blodgett and Drop Cast PbS QD Films Due to Strong Electron–Phonon Coupling. *Nanoscale* **2017**, *9* (43), 17133–17142. https://doi.org/10.1039/C7NR05247B.
(19) Diroll, B. T.; Kirschner, M. S.; Guo, P.; Schaller, R. D. Optical and Physical Probing of Thermal Processes in Semiconductor and Plasmonic Nanocrystals. *Annu. Rev. Phys. Chem.* **2019**, *70* (1), 353–377. https://doi.org/10.1146/annurev-physchem-042018-052639.
(20) Greffet, J.-J.; Bouchon, P.; Brucoli, G.; Marquier, F. Light Emission by Nonequilibrium Bodies: Local Kirchhoff Law. *Phys. Rev. X* **2018**, *8* (2), 021008. https://doi.org/10.1103/PhysRevX.8.021008.
(21) Wang, H.; Aassime, A.; Le Roux, X.; Schilder, N. J.; Greffet, J.-J.; Degiron, A. Revisiting the Role of Metallic Antennas to Control Light Emission by Lead Salt Nanocrystal Assemblies. *Phys. Rev. Appl.* **2018**, *10* (3), 034042. https://doi.org/10.1103/PhysRevApplied.10.034042.
(22) Novotny, L.; Hecht, B. *Principles of Nano-Optics*; 2006.
(23) Konstantatos, G.; Sargent, E. H. PbS Colloidal Quantum Dot Photoconductive Photodetectors: Transport, Traps, and Gain. *Appl. Phys. Lett.* **2007**, *91* (17), 173505. https://doi.org/10.1063/1.2800805.
(24) Voznyy, O.; Levina, L.; Fan, F.; Walters, G.; Fan, J. Z.; Kiani, A.; Ip, A. H.; Thon, S. M.; Proppe, A. H.; Liu, M.; Sargent, E. H. Origins of Stokes Shift in PbS Nanocrystals. *Nano Lett.* **2017**, *17* (12), 7191–7195. https://doi.org/10.1021/acs.nanolett.7b01843.
(25) Lhuillier, E.; Ithurria, S.; Descamps-Mandine, A.; Douillard, T.; Castaing, R.; Xu, X. Z.; Taberna, P.-L.; Simon, P.; Aubin, H.; Dubertret, B. Investigating the N- and p-Type Electrolytic Charging of Colloidal Nanoplatelets. *J. Phys. Chem. C* **2015**, *119* (38), 21795–21799. https://doi.org/10.1021/acs.jpcc.5b05296.
(26) Hu, C.; Gassenq, A.; Justo, Y.; Yakunin, S.; Heiss, W.; Hens, Z.; Roelkens, G. Short-Wave Infrared Colloidal Quantum Dot Photodetectors on Silicon; Razeghi, M., Ed.; San Francisco, California, USA, 2013; p 863127. https://doi.org/10.1117/12.2001246.
(27) Choi, J. J.; Luria, J.; Hyun, B.-R.; Bartnik, A. C.; Sun, L.; Lim, Y.-F.; Marohn, J. A.; Wise, F. W.; Hanrath, T. Photogenerated Exciton Dissociation in Highly Coupled Lead

Salt Nanocrystal Assemblies. *Nano Lett.* **2010**, *10* (5), 1805–1811. https://doi.org/10.1021/nl100498e.

(28) Sun, L.; Choi, J. J.; Stachnik, D.; Bartnik, A. C.; Hyun, B.-R.; Malliaras, G. G.; Hanrath, T.; Wise, F. W. Bright Infrared Quantum-Dot Light-Emitting Diodes through Inter-Dot Spacing Control. *Nat. Nanotechnol.* **2012**, *7* (6), 369–373. https://doi.org/10.1038/nnano.2012.63.

(29) Yazdani, N.; Andermatt, S.; Yarema, M.; Farto, V.; Bani-Hashemian, M. H.; Volk, S.; Lin, W. M. M.; Yarema, O.; Luisier, M.; Wood, V. Charge Transport in Semiconductors Assembled from Nanocrystal Quantum Dots. *Nat. Commun.* **2020**, *11* (1), 2852. https://doi.org/10.1038/s41467-020-16560-7.

(30) Lhuillier, E.; Pedetti, S.; Ithurria, S.; Heuclin, H.; Nadal, B.; Robin, A.; Patriarche, G.; Lequeux, N.; Dubertret, B. Electrolyte-Gated Field Effect Transistor to Probe the Surface Defects and Morphology in Films of Thick CdSe Colloidal Nanoplatelets. *ACS Nano* **2014**, *8* (4), 3813–3820. https://doi.org/10.1021/nn500538n.